\documentclass[a4paper,12pt]{article}
\usepackage[T2A]{fontenc}
\usepackage[cp1251]{inputenc}
\usepackage[english]{babel}
\usepackage{amsmath}
\usepackage{graphicx}
\usepackage{caption2}

\begin{document}
\thispagestyle{empty}

\begin{center}
{\Large \textbf{Correlators of hadron currents: the model and the
ALEPH data on $\tau$-decay.}}

\vspace{2cm}

{\large V. D. Orlovsky}

\vspace{1cm}

\textit{State Research Center \\
Institute of Theoretical and Experimental Physics,
B.~Cheremushkinskaya 25 \\
117218 Moscow, Russia }

\vspace{0.5cm}

\textit{Moscow Engineering Physics Institute (State University),
115409 Moscow, Russia }

\vspace{0.5cm}

e-mail: orlovskii@itep.ru

 \vspace{3cm}

\begin{abstract}
The model with the meson spectrum, consisting of zero-width
equidistant resonances, is considered with connection to current
correlators in coordinate space. The comparison of the explicit
expressions for the correlators, obtained in this model, with the
experimental data of ALEPH collaboration on $\tau$-decay is made
and good agreement is found.
\end{abstract}

\end{center}
\newpage

\section{Introduction}

The main objects of our studies are correlators of charged hadron
currents with the value of isospin $I=1$
\begin{equation}\label{correlator}
P_{\mu\nu}^{(c)}(x)=\langle j_\mu^{(c)}(x) j_\nu^{(c)}(0)^\dag
\rangle
\end{equation}
for vector and axial channels, i.~e. for
\begin{equation}
j_\mu^{(v)}(x)=\bar{u}\gamma_\mu d; \quad j_\mu^{(a)}(x)=
\bar{u}\gamma_\mu\gamma_5d.
\end{equation}

These objects play important role in QCD, since they contain
information of two kinds: one is the spectral features of
hadrons~-- the set of the mass poles $m_n^2$ and the corresponding
quark coupling constants $c_n$, which define behavior of the
correlators for large $x$, on the other hand, at small distances,
where perturbation theory works, and OPE is assumed to be valid,
the coefficients of the correlation functions expansion give us
matrix elements of local field operators~-- the vacuum condensates
like $\langle G_{\mu\nu}^2 \rangle$.

Current correlators were studied long ago in the framework of
perturbation theory, which works at small $x$ because of
asymptotic freedom of QCD \cite{Yndur}. Power corrections to this
behavior, expressed in terms of gluon and quark condensates, were
found with the help of OPE, which is consistent also only for
small enough distances \cite{SVZ}. There are many papers, where
calculations of (\ref{correlator}) are given on the base of
different approaches. So, the random instanton liquid model, where
it is supposed, that the nonperturbative QCD-vacuum is dominated
by strong classical field configuration~-- instantons, whose
position and color orientations are distributed randomly,~-- gives
good result for the difference of the correlation functions in all
region of distances ($0<x<1.5~\mbox{fm}$) \cite{Shuryak}.
Correlation functions and their moments in momentum space are
considered in some papers with the purpose of extraction from them
the values of the lowest OPE condensates \cite{ioffe1},
\cite{ioffe2}, \cite{Narison}. There is another model~-- the
so-called Minimal Hadronic Ansatz (MHA), in which the spectral
density consists of the pion state, a vector state and an axial
state plus continuum. In this this approximation one can compute
corresponding condensates \cite{Friot} or compare the moments of
the polarization operators with the experimental data
\cite{DeRafael}. The method of calculation of correlators, more
general, than MHA, was proposed in the paper \cite{SS}. This
method, as well as MHA, assumes the limit of large number of
colors $N_c$, when the hadronic spectrum is represented by the set
of zero-width levels. One can calculate the spectrum
quasiclassically \cite{spectrum}, restore the correlator in
momentum space by this spectrum and by residues and then go to
coordinate space. In the same paper was made the comparison of
obtained expressions for the correlators with the lattice data
\cite{lat1} and \cite{lat2} and good agreement of the results was
found.

This paper continues the line of research, started in \cite{SS}.
We will be interested in explicit expressions for current
correlators (\ref{correlator}) in the model with equidistant
spectrum, and the main purpose is to compare these expressions
with the real experimental data of ALEPH collaboration
\cite{Aleph} on hadronic $\tau$-decay. Note also, that the model
of zero-width equidistant resonances has been considered with
respect to correlation function in \cite{Golt}, \cite{Beane},
where the main attention was spared to the momentum-space
representation and the corresponding OPE coefficients were
estimated. We will be interested in the current correlators,
calculated in the Euclidean coordinate space and compare them
directly with the experimental curves.

\section{Current correlator: the model and experiment}
We work with QCD in the limit of large number of colors $N_c$. In
the framework of this limit the hadronic spectrum is a set of
zero-width resonances. Masses $m_n$ and residues $c_n$ of current
correlators  are stable at large $N_c$ limit and define the
imaginary part of the polarization operators (see Appendix):
\begin{equation}\label{Im->m_n,c_n}
\frac{1}{\pi}{\rm Im}\Pi(s)=\sum\limits_{n=0}^\infty c_n
\delta(s-m_n^2),
\end{equation}

On the other hand, one can obtain the equidistant mass spectrum
and the set of residues in analytic form for all $n$
quasiclassically, solving the corresponding Hamiltonian problem
for two quarks \cite{SS,spectrum}:
\begin{equation}
(m_n^{(c)})^2=m^2n+m_c^2, \quad c_n=\frac{N_c}{12\pi^2}m^2,
\end{equation}
where $m_c^2$ are the masses of low lying resonances in each
channel, and the quantity $m^2=4\pi\sigma$ is defined universally
for all channels, where $\sigma$ is physical string tension.

In that way one can easily find with the help of
(\ref{Im->m_n,c_n}) the correlation function in the momentum space
and then go to the coordinate one (see Appendix).

It is convenient to consider the normalized correlation functions
and to define the ratio $R^{(c)}(x)$ as
\begin{equation}
R^{(c)}(x)=\frac{P^{(c)}(x)}{P_{free}(x)}
\equiv\frac{g^{\mu\nu}P^{(c)}_{\mu\nu}(x)}{P_{free}(x)},
\end{equation}
where
\begin{equation}\label{P_free}
P_{free}(x)=\frac{6}{\pi^4x^6}.
\end{equation}

Experimental data provide us with information about the spectral
functions $v_1(q^2)$ and $a_1(q^2)$ in the interval $ 0<s<s_0=3.5
\mbox{ GeV}^2$:
\begin{equation}\label{Im->v,a}
{\rm Im} \Pi^{(v)}(s) = \frac{1}{2\pi}v_1(s), \quad {\rm Im}
\Pi_1(s) = \frac{1}{2\pi}a_1(s),
\end{equation}
i.~e. about the imaginary parts of polarization operators
$\Pi^{(v)}(q^2)$ and $\Pi_1(q^2)$, defining the correlation
functions in momentum space (\ref{P_v(q)}) and (\ref{P_a(q)}).

Using the dispersion relation
\begin{equation}
\Pi(q^2)=\int\limits_0^\infty dt \frac{1}{t-q^2}\frac{1}{\pi} {\rm
Im} \Pi(t)
\end{equation}
and the known representation for the McDonald function $K_1(z)$
for $z=\sqrt{-sx^2}$, it is easy to obtain the spectral
representation for correlation functions in coordinate space:
\begin{equation}
P^{(c)}(x_E)= \int\limits_0^\infty s\, \rho^{(c)}(s)D(\sqrt
s,x_E)ds,
\end{equation}
where $x_E=\sqrt{-x^2}$ is a distance in Euclidean space,
$\rho^{(v\setminus a)}(s)=\displaystyle\frac{3}{2\pi^2} \, v_1(s)
\setminus a_1(s)$ are corresponding spectral functions, and
\begin{equation}
D(\sqrt s,x_E)=\frac{\sqrt s}{4\pi^2x_E}K_1(\sqrt sx_E).
\end{equation}

One should take care of the term $q_\mu q_\nu \Pi_2(q^2)$ from
(\ref{P_a(q)}), which corresponds to the contribution of the pion
state. It is taken into account by adding to $P^{(a)}(x_E)$ the
term (see, e.~g. \cite{SS})
\begin{equation}
P_\pi^{(a)}(x_E)= -f_\pi^2m_\pi^2D(m_\pi,x_E).
\end{equation}

The integrals in considered combinations
\begin{equation}
P_{v\pm a}(x_E)= \int\limits_0^\infty s\, \rho_{v\pm a}(s)D(\sqrt
s,x_E)ds \mp f_\pi^2m_\pi^2D(m_\pi,x_E),
\end{equation}
where $\rho_{v\pm a}(s)=\rho^{(v)}(s)\pm\rho^{(a)}(s), \, P_{v\pm
a}=P^{(v)}\pm P^{(a)}$, were taken numerically by Simpson method,
using the experimental data for difference and sum of the spectral
functions in the region $0<s<s_0$. The spectral functions
$\rho_{v\pm a}(s)$ were replaced by the parton model prediction
for $s>s_0$, i.~e. $v_1(s)+a_1(s)=1, \quad v_1(s)-a_1(s)=0$ for
$s>s_0$. Under this replacement we do not take into account
perturbation theory corrections, which change little behavior of
the sum of the current correlators for small $x$. However, this
perturbative corrections, corresponding to the hybrid states in
the spectral sum in terms of our model, are not taken into account
also, so the disregarding of them in the experimental curves is
coordinated with the model calculations.

\begin{figure}[p]
  \renewcommand{\captionlabeldelim}{.}
  \renewcommand{\figurename}{Fig.}
  \begin{center}
   \includegraphics[height=8.5cm,width=8.5cm]{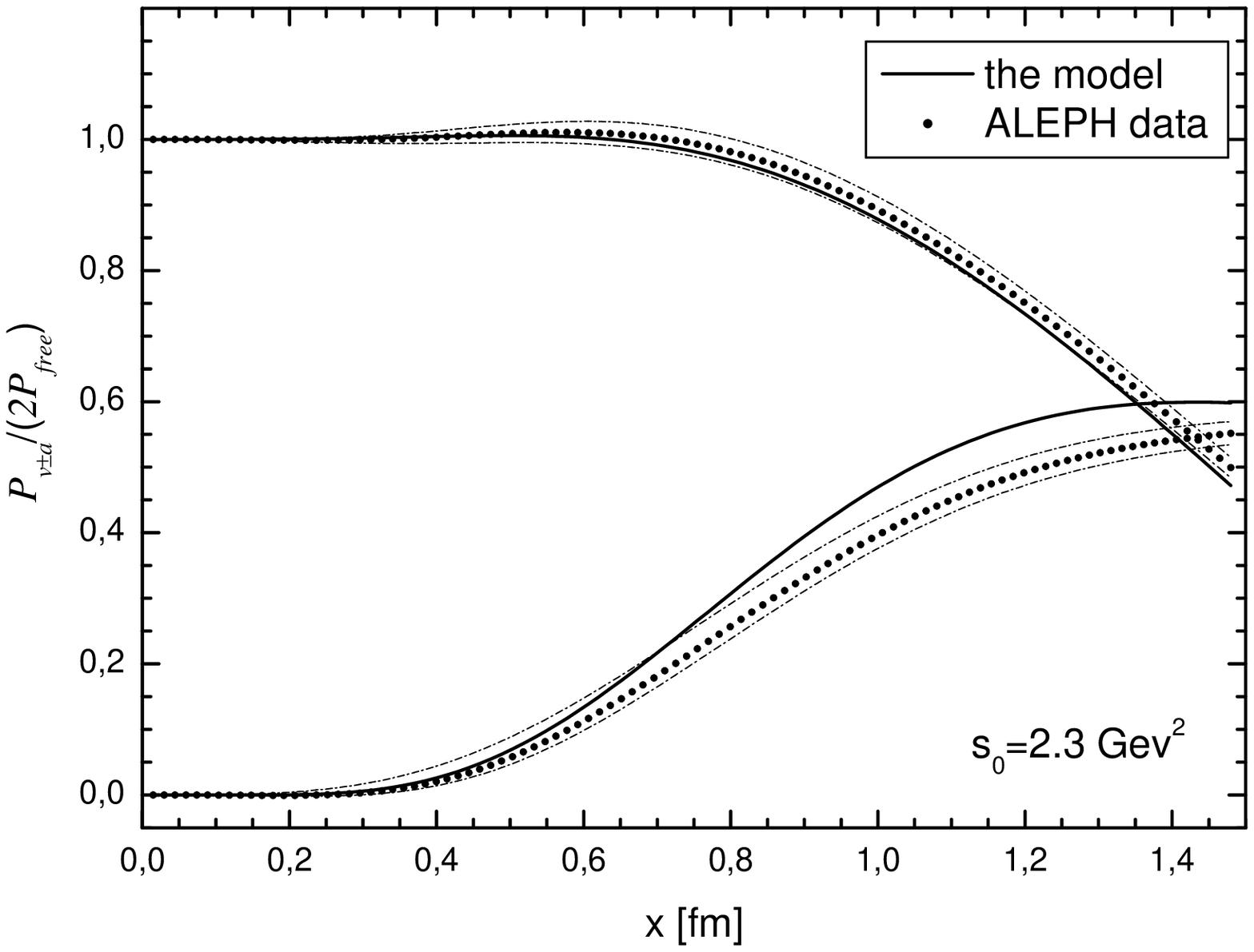}
   \includegraphics[height=8.5cm,width=8.5cm]{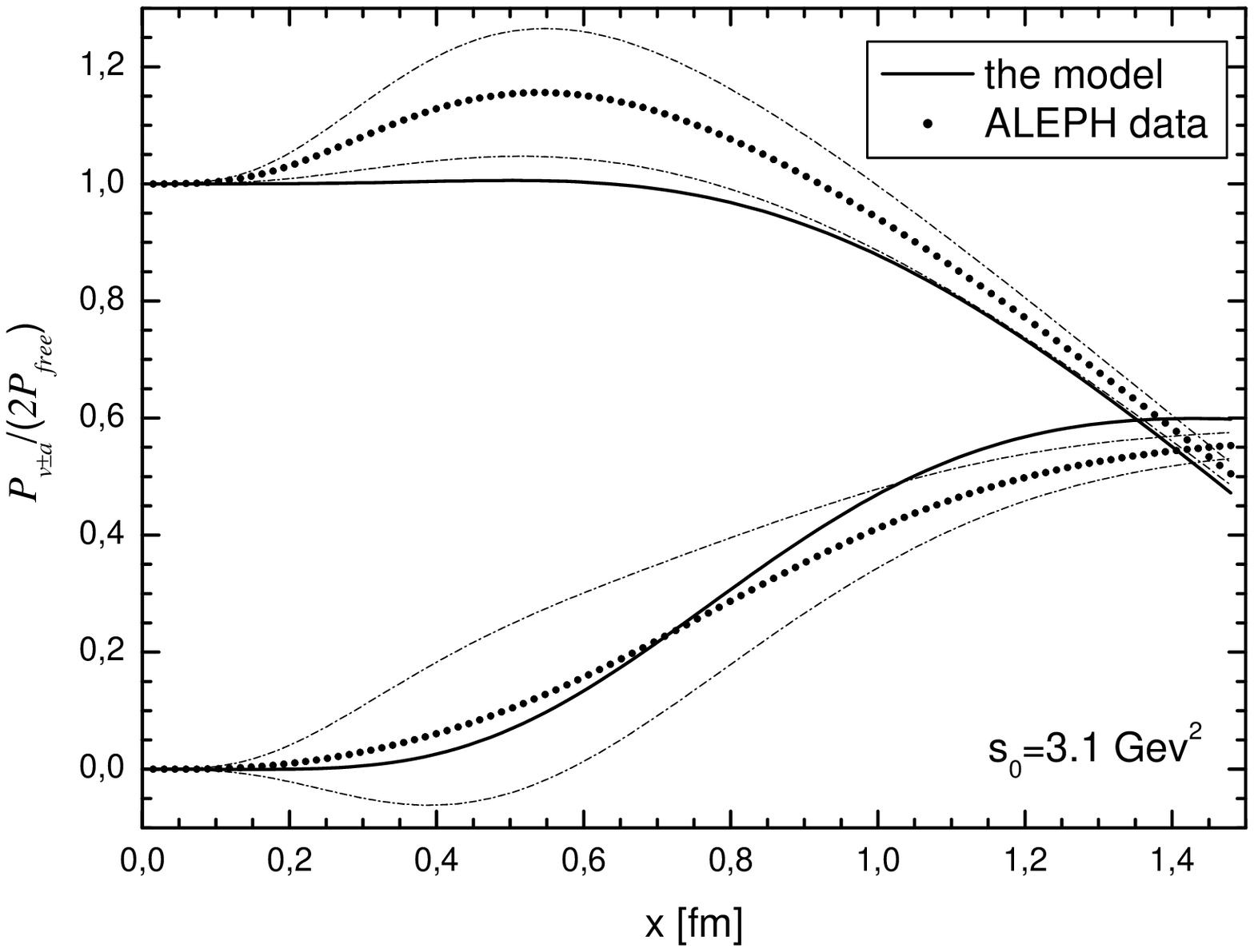}
  \end{center}
  \caption{Normalized Euclidean coordinate space correlation functions $P_{v\pm a}(x)$ for $s_0=2.3 \mbox{ GeV}^2$ and $s_0=3.1 \mbox{ GeV}^2$.}\label{result1}
 \end{figure}

The experimental curves, normalized to the free answer
(\ref{P_free}), and the corresponding model predictions are shown
on the Fig.~\ref{result1} for two values of the cut-off scale
$s_0=2.3 \mbox{ GeV}^2$ and $s_0=3.1 \mbox{ GeV}^2$. Notice, that
the choice of the cut-off parameter $s_0$ changes noticeably
behavior of $P_{v+a}(x_E)$ only in the intermediate region of $x$,
pushing away the theoretical curve from the experimental one. In
addition, the error bars increase with increasing $s_0$ in this
region because of large uncertainties of the experimental data for
$s_0>2.8 \mbox{ GeV}^2$, so agreement of the experimental curves
with the model predictions remains acceptable even on the
quantitative level.

\begin{figure}[p]
  \renewcommand{\captionlabeldelim}{.}
    \renewcommand{\figurename}{Fig.}
   \begin{center}
   \includegraphics[width=10cm]{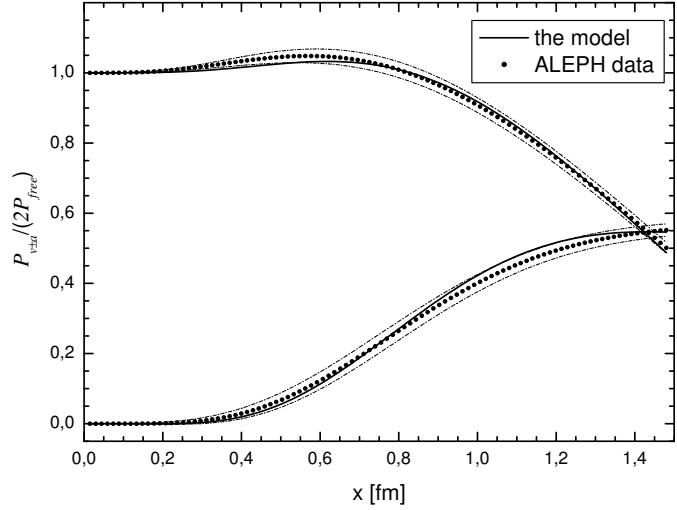}
   \end{center}
  \caption{Normalized Euclidean coordinate space correlation functions $P_{v\pm a}(x)$ for the set of parameters (13) and $s_0=2.5 \mbox{ GeV}^2$.}\label{result2}
 \end{figure}

On the other hand, using meson masses and decay constants as
fitting parameters, one can place the curves, obtained in the
framework of the model at large $N_c$, inside the error corridors
completely (see Fig.~\ref{result2}). The corresponding values
\begin{gather}\label{constants2}
m_\rho^2=0.62 \mbox{ GeV}^2, \quad \lambda_\rho^2=0.048 \mbox{ GeV}^2, \notag \\
m_\pi=135 \mbox{ MeV},  \quad f_\pi=89 \mbox{ MeV}, \\
m_{a_1}=1.18 \mbox{ GeV} \notag
\end{gather}
reproduce the table values (21) with $2-4\%$ accuracy.

The errors was computed with the help of experimental covariance
matrices
\begin{equation}
\rho_{v\pm a}(s,s') = \langle \Delta\rho_{v\pm a}(s)\,
\Delta\rho_{v\pm a}(s') \rangle
\end{equation}
by the standard expression
\begin{equation}
\Delta P_{v\pm a}(x_E)= [\langle P_{v\pm a}^2\rangle-\langle
P_{v\pm a} \rangle^2]^{1/2} = \left[ \int\limits_0^{s_0} ds
\int\limits_0^{s_0} ds' \, \rho_{v\pm a}(s,s') \, s\, D(\sqrt
s,x_E) \, s' \, D(\sqrt{s'},x_E) \right]^{1/2}.
\end{equation}

\section{Conclusion}
We have considered the model, based on QCD in the limit of large
number of colors $N_c$, in which the meson spectrum is taken as
the series of zero-width equidistant levels. This model is surely
far from the real physical world and is over simplified. However,
as can be seen from the results of comparison of the calculations,
performed in the framework of this model, with the high precision
experimental data of ALEPH collaboration for the correlators of
hadron currents, the knowledge of several spectral features (in
our case, these are the data about low lying resonances in
corresponding channels and the correct behavior of the mass
spectrum and of the set of residues for $n\rightarrow\infty$) is
enough to reproduce correctly on qualitative and even on
quantitative level such integral objects (in a sense of the
spectral density) as correlation functions (\ref{correlator}).
Besides, note also, that all the parameters of the model (meson
masses, decay constants, string tension) are fixed by their
physical values, and calculated correlation function have in
principle a form of a fixed prediction without any fitting
parameters, however, having shifted quite a little the values of
this input parameters ($\sim2-3\%$), one can reach a very good
agreement with the experimental data.

\section*{Acknowledgments}
I am grateful to Dr. V.~I.~Shevchenko for numerous discussions of
the problems, considered in the paper. I wish to thank Prof.
Yu.~A.~Simonov and Dr. K.~N.~Zyablyuk for useful discussions. I am
also thankful to the "Dynasty" \, foundation for financial
support.

\newpage

\section*{Appendix}
For conserved vector current one can write the following
expression in momentum space:
\begin{equation}\label{P_v(q)}
P_{\mu\nu}^{(v)}(q) = i\int d^4x P_{\mu\nu}^{(v)}(x) \exp(iqx) =
(q_\mu q_\nu-q^2g_{\mu\nu})\Pi^{(v)}(q^2).
\end{equation}

There are two Lorentz structure in the axial channel:
\begin{equation}\label{P_a(q)}
P_{\mu\nu}^{(a)}(q) = i\int d^4x P_{\mu\nu}^{(a)}(x) \exp(iqx) =
(q_\mu q_\nu-q^2g_{\mu\nu})\Pi_1(q^2)+ q_\mu q_\nu \Pi_2(q^2).
\end{equation}

The normalized correlation functions in the Euclidean coordinate
space are given by following expressions in accordance with result
of \cite{SS}:
\begin{equation}
R^{(v)}(z_v)=\xi z_v^5 K_1(z_v) + \frac{b_vz_v^6}{2^7} \,
\bar{p}(b_v,z_v)
\end{equation}

\begin{equation}
R^{(a)}(z_a)=-\frac{\pi^2}{12}f^2\gamma^3z_a^5K_1(\gamma z_a) +
\frac{b_az_a^6}{2^7}\, \bar{p}(b_a,z_a),
\end{equation}
which are depend on the dimensionless coordinates
$z_c=\sqrt{-x^2}m_c$, where $m_c$ are the masses of the resonances
in vector ($\rho$~-~meson) and axial ($a_1$~-~meson) channels. The
contributions of the pion and the $\rho$~-~meson states are taken
into account separately and are described by the first members in
$R^{(c)}(x)$. The parameters
\begin{equation}
b_c=\frac{4\pi\sigma}{m_c^2}, \quad \xi=\frac{\pi^2}{8}
\left(\frac{\lambda_\rho^2}{m_v^2} - \frac{b_v}{4\pi^2} \right),
\quad \gamma=\frac{m_\pi}{m_a}, \quad f=\frac{f_\pi}{m_a}
\end{equation}
are fixed by the physical values of meson masses, decay constants
and string tension constant:
\begin{gather}\label{cnst}
m_\rho^2=0.6 \mbox{ GeV}^2, \quad \lambda_\rho^2=0.047 \mbox{ GeV}^2, \notag \\
m_\pi=140 \mbox{ MeV},  \quad f_\pi=92 \mbox{ MeV}, \\
m_{a_1}=1.23 \mbox{ GeV}, \quad \sigma=0.17 \mbox{ GeV}^2 \notag.
\end{gather}

Remaining terms, containing the function
\begin{equation}
\bar{p}(b,z)=\int\limits_0^\infty \frac{du}{u^2} \exp \left(
-\frac{z^2}{4u}-u \right)
\frac{1+\exp(-bu)(b-1)}{(1-\exp(-bu))^2},
\end{equation}
are determined by contribution of a set of the equidistant levels
with the masses
\begin{equation}
(m_n^{(c)})^2=m^2n+m_c^2, \quad m^2=4\pi\sigma.
\end{equation}
\newpage

\end{document}